\documentclass[runningheads]{cl2emult}

\usepackage{makeidx}  
\usepackage{graphicx} 
\usepackage{multicol} 
\makeindex            

\newcommand{\be}{\begin{equation}}
\newcommand{\ee}{\end{equation}}
\newcommand{\bea}{\begin{eqnarray}}
\newcommand{\eea}{\end{eqnarray}}
\newcommand{\lb}{\label}

\newcommand{\I}{\mbox{i}}
\newcommand{\E}{\mbox{e}}
\newcommand{\D}{\mbox{d}}

\begin{document}
\title*{Conceptual issues in quantum cosmology\thanks{Report
         Freiburg THEP-99/7. To appear in the Proceedings of the
        Karpacz Winter School on {\em From Cosmology to Quantum Gravity} (Springer~1999).}} 
\toctitle{Conceptual issues in quantum cosmology}
%
%
\titlerunning{Quantum cosmology}
%
\author{Claus Kiefer\inst{1,2}}
\authorrunning{Claus Kiefer}
%
%
\institute{Fakult\"at f\"ur Physik, Universit\"at Freiburg,
           Hermann-Herder-Str.~3, D-79104~Freiburg, Germany
\and Institute for Advanced Study Berlin, Wallotstr.~19,
     D-14193 Berlin, Germany}

\maketitle              

\begin{abstract}
I give a review of the conceptual issues that arise in
theories of quantum cosmology. I start by emphasising some
features of ordinary quantum theory that also play a crucial
role in understanding quantum cosmology. I then give motivations
why spacetime cannot be treated classically at the most fundamental
level. Two important issues in quantum cosmology -- the problem
of time and the role of boundary conditions -- are discussed
at some length. Finally, I discuss how classical spacetime
can be recovered as an approximate notion. This involves the
application of a semiclassical approximation and the process
of decoherence. The latter is applied to both global degrees
of freedom and primordial fluctuations in an inflationary
Universe.
\end{abstract}

\section{Introduction}
As the title of this school indicates, a consistent quantum theory
of gravity is eventually needed to solve the fundamental cosmological
questions. These concern in particular the role of initial
conditions and a deeper understanding of processes such as inflation.
The presence of the singularity theorems in general relativity
prevents the formulation of viable initial conditions in the classical
theory. Moreover, the inflationary scenario can be successfully
implemented only if the cosmological no-hair conjecture is
imposed -- a conjecture which heavily relies on assumptions
about the physics at sub-Planckian scales. 

It is generally assumed that a quantum theory of gravity can
cure these problems. This is not a logical necessity, though,
since there might exist classical theories which could achieve
the same. As will be discussed in my contribution, however,
one can put forward many arguments in favour of the quantisation
of gravity, which is why classical alternatives will not be considered
here.

Although a final quantum theory of gravity is still elusive,
there exist concrete approaches which are mature enough
to discuss their impact on cosmology. Here I shall focus on
conceptual, rather than technical, issues that one might expect
to play a role in any quantum theory of the gravitational field.
In fact, most of the existing approaches leave the basic structures
of quantum theory, such as its linearity, untouched.

Two aspects of quantum cosmology must be distinguished.
The first is concerned with the application of quantum theory
to the Universe as a whole and is independent of any particular
interaction. This raises such issues as the interpretation 
of quantum theory for closed systems, where no external
measuring agency can be assumed to exist. In particular,
it must be clarified how and to what extent classical properties
emerge. The second aspect deals with the peculiarities 
that enter through quantum aspects of the gravitational interaction.
Since gravity is the dominant interaction on the largest scales,
this is an important issue in cosmology. Both aspects will be
discussed in my contribution.

Since many features in quantum cosmology arise from
the application of standard quantum theory to the Universe
as a whole, I shall start in the next section with a
dicussion of the lessons that can be learnt from
 ordinary quantum theory. In particular, the 
central issue of the quantum-to-classical transition
will be discussed at some length. Section~3 is then devoted
to full quantum cosmology: I start with giving precise
arguments why one must expect that the gravitational field
is of a quantum nature at the most fundamental level. I then
discuss the problem of time and related issues such as the
Hilbert-space problem. I also
devote some space to the central question of how to 
impose boundary conditions properly in quantum cosmology. 
The last section will then be concerned with the emergence
of a classical Universe from quantum cosmology. 
I demonstrate how an approximate notion of
a time parameter can be recovered
from ``timeless'' quantum cosmology  through some semiclassical
approximation. I then discuss at length the emergence of
a classical spacetime by decoherence. This is important for both
the consistency of the inflationary scenario as well
as for the classicality of primordial fluctuations which
can serve as seeds for galaxy formation and which can be
observed in the anisotropy spectrum of the cosmic
microwave background.

\section{Lessons from quantum theory}
\subsection{Superposition principle and ``measurements''}
The superposition principle lies at the heart of quantum theory.
{}From a conceptual point of view, it is appropriate to separate
it into a kinematical and a dynamical version (Giulini et al.~1996):
\begin{itemize}
\item {\em Kinematical version}: If $\Psi_1$ and $\Psi_2$ physical
states, then $\alpha\Psi_1+\beta\Psi_2$, where $\alpha$ and
$\beta$ are complex numbers, is again a physical state.
\item {\em Dynamical version}: If $\Psi_1(t)$ and $\Psi_2(t)$
are solutions of the Schr\"odinger equation, then
$\alpha\Psi_1(t)+\beta\Psi_2(t)$ is again a solution of the
Schr\"odinger equation. 
\end{itemize}
These features give rise to the {\em nonseparability} of quantum
theory. If interactions between systems are present, the
emergence of entangled states is unavoidable. As
Schr\"odinger (1935) put it: 
\begin{quote}
I would not call that {\em one} but rather {\em the}
characteristic trait of quantum mechanics, the one that enforces
its entire departure from classical lines of thought. By the
interaction the two representatives (or $\psi$-functions)
have become entangled. \ldots Another way of expressing the
peculiar situation is: the best possible knowledge of a
{\em whole} does not necessarily include the best possible
knowledge of all its {\em parts}, even though they may be
entirely separated \ldots
\end{quote}
Because of the superposition principle, quantum states
which mimic classical states (for example, by being localised),
form only a tiny subset of all possible states. Up to now,
no violation of the superposition principle has been observed
in quantum-mechanical experiments, and the only question is
why we observe classical states at all. After all, one would
expect the superposition principle to have unrestricted
validity, since also macroscopic objects are composed
of atoms.

The power of the superposition principle was already
noted by von Neumann in 1932 when he tried to describe
the measurement process consistently in quantum terms. 
He considers an interaction between a system and a (macroscopic)
apparatus (cf. Giulini et al.~1996).
Let the states of the measured system which are
discriminated by the apparatus be denoted by $|n\rangle$, then an
appropriate interaction Hamiltonian has the form
\be H_{int} =\sum_n|n\rangle\langle n| \otimes\hat{A}_n\ . \ee
The operators $\hat{A}_n$, acting on the states of the apparatus, are
rather
arbitrary, but must of course depend on the ``quantum number" $n$.
Note that the measured ``observable" is dynamically defined by
the system-apparatus interaction and there is no reason to introduce
it
axiomatically (or as an additional concept).
If the
measured system is initially in the state $|n\rangle$ and the device
 in some initial state $|\Phi_0\rangle$,
the evolution according to the Schr\"odinger equation
with Hamiltonian (1) reads
\bea |n\rangle|\Phi_0\rangle \stackrel{t}{\longrightarrow}
     \exp\left(-\I H_{int}t\right)|n\rangle|\Phi_0\rangle
     &=& |n\rangle\exp\left(-\I \hat{A}_nt\right)|\Phi_0\rangle\nonumber
\\
     &=:& |n\rangle|\Phi_n(t)\rangle\ . \eea
The resulting apparatus states $|\Phi_n(t)\rangle$ are usually called
``pointer positions''.
An analogy to (2) can also be
written down in classical physics. The essential new quantum features
come into play when we consider a {\em superposition} of different
eigenstates (of the measured ``observable'') as initial state. The
linearity of time evolution immediately leads to
\be \left(\sum_n c_n|n\rangle\right)|\Phi_0\rangle
    \stackrel{t}\longrightarrow\sum_n c_n|n\rangle
    |\Phi_n(t)\rangle\ . \lb{sup} \ee
This state does not, however, correspond to a definite
measurement result -- it contains a ``weird'' superposition
of macroscopic pointer positions! This motivated von Neumann
to introduce a ``collapse'' of the wave function, because he
saw no other possibility to adapt the formalism to experience.
There have been only rather recently attempts to 
give a concrete dynamical formulation of this collapse
(see, e.g., Chap.~8 in Giulini et al. (1996)). However, none
of these collapse models has yet been experimentally confirmed.
In the following I shall review a concept that enables one
to reconcile quantum theory with experience
without introducing an explicit collapse; strangely enough, it is
the superposition principle itself that leads to
classical properties.  
 
\subsection{Decoherence: Concepts, examples, experiments}
The crucial observation is that macroscopic objects
cannot be considered as being isolated -- they are 
unavoidably coupled to ubiquitous degrees of freedom of
their einvironment, leading to quantum entanglement.
 As will be briefly discussed in the
course of this subsection, this gives rise to classical
properties for such objects -- a process known
as {\em decoherence}. This was first discussed by Zeh in the
seventies and later elaborated by many authors;
a comprehensive treatment is given by Giulini et al. (1996),
other reviews include Zurek (1991), Kiefer and Joos (1999),
see also the contributions to the volume Blanchard et al. (1999).

Denoting the environmental states with $\vert{\cal E}_n\rangle$,
the interaction with system and apparatus yields instead of
(\ref{sup}) a superposition of the type
\be
\left(\sum_n c_n|n\rangle\right)|\Phi_0\rangle|{\cal E}_0\rangle
    \stackrel{t}\longrightarrow\sum_n c_n|n\rangle
    |\Phi_n\rangle|{\cal E}_n\rangle\ . \lb{supe} \ee
This is again a macroscopic superposition, involving 
a tremendous number of degrees of freedom. The crucial point now is,
however, that most of the environmental degrees of freedom 
are not amenable to observation. If we ask what can be seen
when observing only system and apparatus, we need --
according to the quantum rules -- to calculate the reduced density
matrix $\rho$ that is obtained from (\ref{supe}) upon tracing
out the environmental degrees of freedom. 
  
If the environmental states are approximately orthogonal
(which is the generic case),
\be \langle{\cal E}_m|{\cal E}_n\rangle \approx \delta_{mn}\ , \ee
the density
matrix becomes approximately diagonal in the ``pointer basis'',
\be \rho_S \approx \sum_n|c_n|^2|n\rangle\langle n|
       \otimes |\Phi_n\rangle\langle\Phi_n|
       \ . \ee
Thus, the result of this interaction is a density matrix which seems
to describe an ensemble of different outcomes $n$ with the
respective probabilities. One must be careful in analysing its
interpretation,
however: This density matrix only corresponds to an {\it apparent}
ensemble,
not a genuine ensemble of quantum states.
What can safely be stated is the fact, that interference terms 
(non-diagonal elements) are absent locally, although
they are still present in the total system, see (\ref{supe}).
The coherence present in the initial system state in (3) can no longer
 be observed; 
it is {\it delocalised} into the larger system. As is well known, any
interpretation of a superposition as an ensemble
of components can be disproved experimentally
by creating interference effects. The same is true for 
the situation described in (3). For
example, the evolution could {\it in principle} be reversed. Needless
to say that such a reversal is experimentally extremely difficult, but
the interpretation and consistency of a physical theory must not depend
on our present technical abilities. Nevertheless, one often finds
explicit
or implicit statements to the effect that the above processes are 
equivalent to the collapse of the wave function (or even solve
the measurement problem). Such statements are certainly unfounded. What
can safely be said, is that coherence between the subspaces of the
Hilbert space spanned by $|n\rangle$ can no longer be observed
in the system considered, {\it if} the
process described by (3) is practically irreversible.

The essential implications are twofold: First, processes of the kind
(3) do happen
frequently and unavoidably for all macroscopic objects. Second, these
processes are irreversible in practically all
realistic situtations. In a normal measurement process, the
interaction and the state of the apparatus are controllable to some
extent (for example, the initial state of the apparatus is known to the
experimenter). In the case of decoherence, typically the initial state
is not known in detail (a standard example is interaction with
thermal radiation), but the consequences for the local density
matrix are the same: If the environment is described by an ensemble,
each member of this ensemble can act in the way described above.

A complete treatment of realistic cases has to include the Hamiltonian
governing the evolution of the system itself (as well as that of the
environment). The exact dynamics of a subsystem is hardly manageable
 (formally it is given by a complicated integro-differential
equation, see Chap.~7 of Giulini et al. 1996).
Nevertheless, we can find important approximate solutions
in some simplifying cases. One example is concerned with
localisation through scattering processes
 and will be briefly discussed in the following.
My treatment will closely follow Kiefer and Joos (1999).

Why do macroscopic objects always appear localised in space? Coherence 
between macroscopically different positions is destroyed {\it very}
rapidly because
of the strong influence of scattering processes. The formal description
may proceed as follows. Let $|x\rangle$ be the position eigenstate
of a macroscopic object, and $|\chi\rangle$ the state of the
incoming particle.
 Following the von Neumann scheme, the scattering of
such particles off an object located at position $x$ may be written as
\be |x\rangle|\chi\rangle \stackrel{t}{\longrightarrow}
    |x\rangle|\chi_x\rangle=|x\rangle S_x|\chi\rangle\ , \ee
where the scattered state may conveniently be calculated by means of
an appropriate S-matrix. For the more general initial state of a wave
packet we have then
\be \int\D^3x\ \varphi(x)|x\rangle|\chi\rangle
    \stackrel{t}{\longrightarrow}\int\D^3x\ \varphi(x)|x\rangle
     S_x|\chi\rangle\ , \ee
and the reduced density matrix describing our object changes
into
\be \rho(x,x')=\varphi(x)\varphi^*(x')
  \left\langle\chi|S_{x'}^{\dagger}S_x|\chi\right\rangle\ .
  \lb{trace} \ee
These steps correspond to the general steps discussed above.
Of course, a single scattering process will usually not resolve a small
distance, so
in most cases the matrix element on the right-hand side
 of (\ref{trace}) will be close to unity.
But if we add the contributions of many scattering processes, an
exponential damping of spatial coherence results:
\be \rho(x,x',t)= \rho(x,x',0)\exp\left\{-\Lambda t(x-x')^2\right\}\ .
\lb{dm} \ee
The strength of this effect is described by a single parameter $\Lambda$
which may be called the ``localisation rate" and is given by
\be \Lambda= \frac{k^2Nv\sigma_{eff}}{V}\ .\ee
Here, $k$ is the wave number of the incoming particles, $Nv/V$
the flux, and $\sigma_{eff}$ is of the order of the total cross section
(for details see Joos and Zeh 1985 or
Sect. 3.2.1 and Appendix 1 in Giulini et al. 1996). 
Some values of $\Lambda$ are given in the Table.

\begin{table}[htb]
\caption[ ]{Localisation rate $\Lambda$ in $\mbox{cm}^{-2}
\mbox{s}^{-1}$ for three sizes of ``dust particles" and various
types of scattering processes (from Joos and Zeh~1985).
This quantity measures how fast interference between different
positions disappears as a function of distance in the course of
time, see (\ref{dm}).}
\begin{flushleft}
\renewcommand{\arraystretch}{1.2}  
\begin{tabular}{l|lll}
\hline
{} & \ $a=10^{-3}\mbox{cm}$ &\  $a=10^{-5}\mbox{cm}$ &\
$a=10^{-6}\mbox{cm}$\\
{} & \ dust particle        &\  dust particle        &\ large molecule
\\ \hline
Cosmic background radiation &\ $10^{6}$& $10^{-6}$ &$10^{-12}$ \\
300 K photons &\ $10^{19}$ & $10^{12}$ & $10^6$ \\
Sunlight (on earth) &\ $10^{21}$  & $10^{17}$ & $10^{13}$ \\
Air molecules & \  $10^{36}$ & $10^{32}$ & $10^{30}$ \\
Laboratory vacuum & \ $10^{23}$ & $10^{19}$ & $10^{17}$\\
($10^3$ particles/$\mbox{cm}^3$) & {}&{}&\\
\hline
\end{tabular}
\renewcommand{\arraystretch}{1}
\end{flushleft}\end{table}

Most of the numbers in the table are quite large, showing the extremely
strong coupling of macroscopic objects, such as dust particles, to their
natural environment. Even in intergalactic space, the 3K background
radiation cannot be neglected.

In a general treatment one must combine the decohering influence
of scattering processes with the internal dynamics of the system.
This leads to master equations for the reduced density matrix,
which can be solved explicitly in simple cases. Let me mention
the example where the internal dynamics is given by the
free Hamiltonian and consider the coherence length, i.e.
the non-diagonal part of the density matrix.
According to the Schr\"odinger equation, a free wave packet would
spread,
thereby increasing its size and extending its coherence properties over
a larger region of space. Decoherence is expected to counteract this
behaviour and reduce the coherence length. This can be seen in the
solution shown in Fig. 1, where the time dependence of the coherence
length (the width of the density matrix in the off-diagonal direction)
is
plotted for a truly free particle
 (obeying a Schr\"odinger equation) and also for increasing strength of
decoherence. For large times the spreading of the
wave packet no longer occurs and the coherence length always decreases
proportional to $1/\sqrt{\Lambda t}$. More details and
more complicated examples can be found in Giulini et al. (1996).

\begin{figure}
\begin{center}
\includegraphics[width=1.0\textwidth]{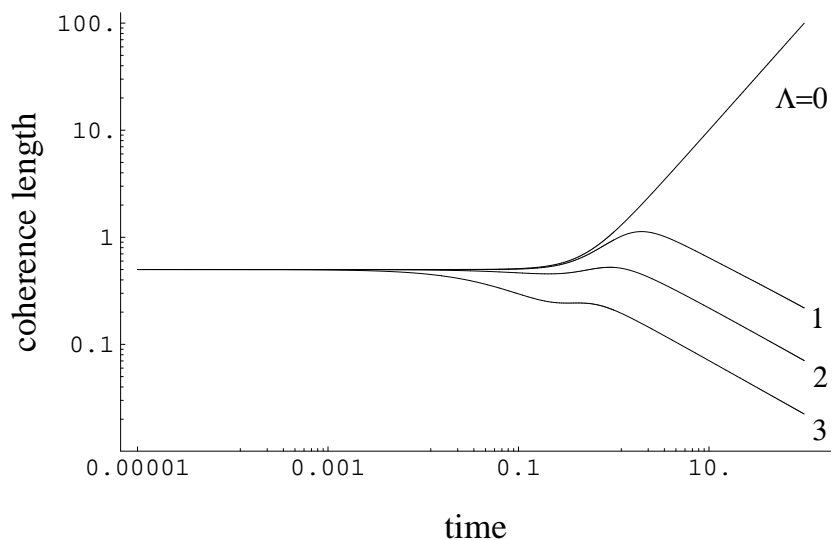}
\end{center}
\caption[]{Time dependence of coherence length. It is a measure of
the spatial extension over which the object can show interference
effects. Except for zero coupling ($\Lambda=0$), the coherence
length always decreases for large times. From Giulini et al. (1996).}
\end{figure}

Not only the centre-of-mass position of dust particles becomes
``classical'' via 
decoherence. The spatial structure of molecules represents another most
important example. Consider a simple model of a chiral molecule (Fig.
2).

\begin{figure}
\begin{center}
\includegraphics[width=1.0\textwidth]{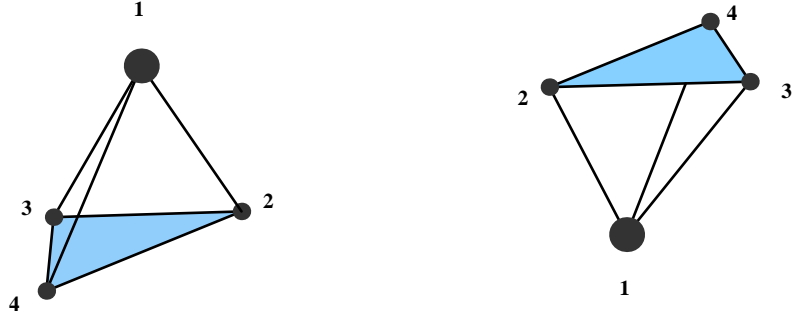}
\end{center}
\caption[]{Typical structure of an optically active, chiral molecule.
Both versions are mirror-images of each other and are not connected
by a proper rotation, if the four elements are different.}
\end{figure}

Right- and left-handed versions both have a rather well-defined spatial
structure, whereas the ground state is - for symmetry reasons -
a superposition of both chiral states. These chiral configurations are
usually separated by a tunneling barrier (compare Fig. 3) which is so
high that under
normal circumstances tunneling is very improbable, as was already
shown by Hund in 1929. But this alone does not explain why chiral
molecules are never found in energy eigenstates! Only the
interaction with the environment can lead to the localisation
and the emergence of a spatial structure. We shall encounter
a similar case of ``symmetry breaking'' in the case of
quantum cosmology, see Sect.~4.2 below.

\begin{figure}
\begin{center}
\includegraphics[width=1.0\textwidth]{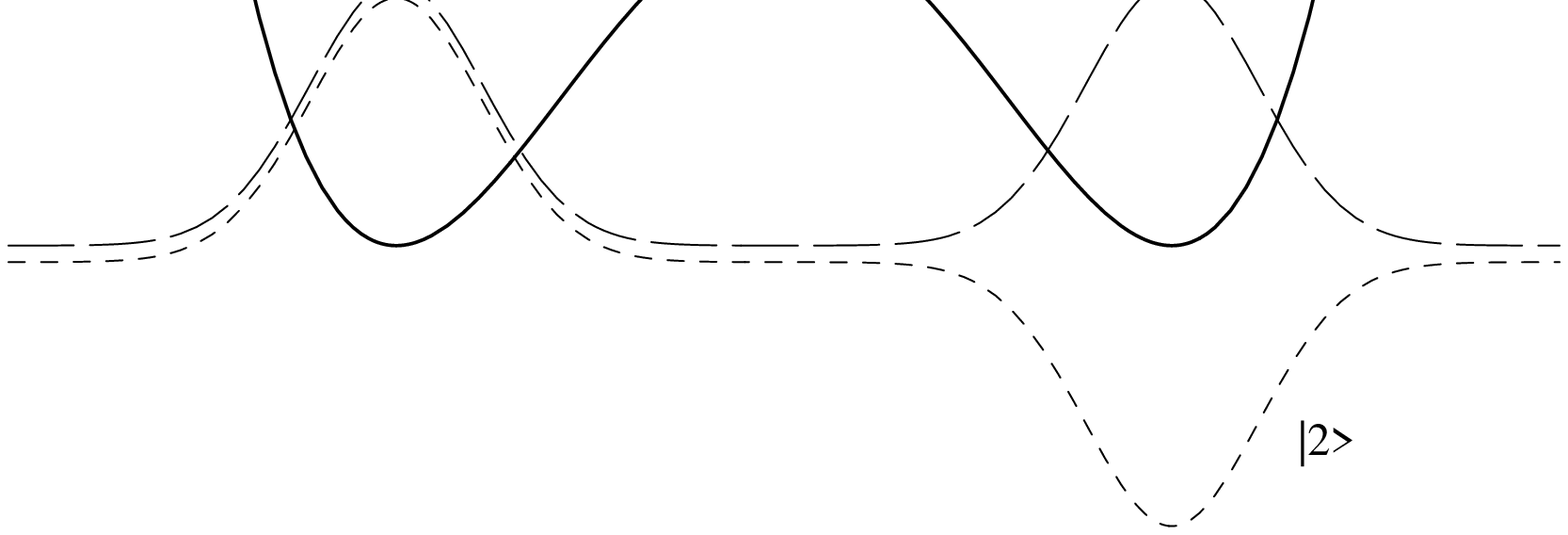}
\end{center}
\caption[]{Effective potential for the inversion coordinate in a model
for a chiral molecule and the two lowest-lying eigenstates. The
ground state is symmetrically distributed over the two wells. Only
linear combinations of the two lowest-lying states are localised and
correspond to a classical configuration.}
\end{figure}

I want to emphasise that decoherence should not be confused
with thermalisation, although they sometimes occur together.
In general, decoherence and relaxation have drastically different
timescales -- for a typical macroscopic situation decoherence
is faster by forty orders of magnitude.
This short decoherence timescale leads to the impression
of discontinuities, e.g. ``quantum jumps'',
although the underlying dynamics, the Schr\"odinger equation,
is continuous. Therefore, to come up with a precise
experimental test of decoherence, one must spend 
considerable effort to bring the decoherence timescale into
a regime where it is comparable with other timescales of
the system. This was achieved by a quantum-optical experiment
that was performed in Paris in 1996, see Haroche (1998) for a review.

What is done in this experiment? The role of the system is
played by a rubidium atom and its states $|n\rangle$ are two
Rydberg states $|+\rangle$ and $|-\rangle$. This atom is sent into a 
high-Q cavity and brought into interaction with an
electromagnetic field. This field plays the role of the
``apparatus'' and its pointer states $|\Phi_n\rangle$ are coherent
states $|\alpha_+\rangle$ and $|\alpha_-\rangle$ which are
correlated with the system states $|+\rangle$ and
$|-\rangle$, respectively. The atom is brought into a superposition
of $|+\rangle$ and $|-\rangle$ which it imparts on the
coherent states of the electromagnetic field; the latter is
then in a superposition of $|\alpha_+\rangle$
 and $|\alpha_-\rangle$, which
resembles a Schr\"odinger-cat state. The role of the
environment is played by mirror defects and the corresponding
environmental states are correlated with the respective
components of the field superposition. One would thus expect
that decoherence turns this superposition locally into a
mixture. The decoherence time is calculated to be
$t_D\approx t_R/\bar{n}$, where $t_R$ is the relaxation time
(the field-energy decay time) and $\bar{n}$ is the average photon number
in the cavity. In the experiment $t_R$ is about 160 microseconds,
and $\bar{n}\approx 3.3$. These values enable one to
monitor the process of decoherence as a process in time.

The decay of field coherence is measured
by sending a second atom with different
delay times into the cavity, playing the
role of a ``quantum mouse''; interference fringes are observed
through two-atom correlation signals. The experimental results
are found to be in complete agreement with the theoretical
prediction. If a value of $\bar{n}\approx 10$ is chosen,
decoherence is already so rapid that no coherence can be seen.
This makes it obvious why decoherence for
macroscopic objects happens ``instantaneously'' for all practical purposes.  

\subsection{On the interpretation of quantum theory\protect
           \footnote{This is adapted from Sect.~4 of
                     Kiefer and Joos (1999).}}
It would have been possible to study
the emergence of classical properties by decoherence
 already in the early
days of quantum mechanics and, in fact, the contributions of
Landau, Mott, and Heisenberg at the end of the twenties can be
interpreted
as a first step in this direction. Why did one not go further
at that time? One major reason was certainly the advent of the
``Copenhagen doctrine'' that was sufficient to apply the formalism
of quantum theory on a pragmatic level. In addition, the imagination
that objects can be isolated from their environment was so deeply
rooted since the time of Galileo, that the {\em quantitative} aspect
of decoherence was largely underestimated. This quantitative
aspect was only borne out by detailed calculations, some of which
I have reviewed above. Moreover, direct experimental verification
was only possible quite recently.

What are the achievements of the decoherence mechanism?
Decoherence can certainly explain why and how {\em within}
quantum theory certain objects (including fields) {\em appear}
classical to ``local'' observers. It can, of course, not explain
why there are such local observers at all. The classical properties
are defined by the {\em pointer basis} for the object,
which is distinguished by the interaction with the environment
and which is sufficiently stable in time. It is important to
emphasise that classical properties are {\em not} an a priori
attribute of objects, but only come into being through the
interaction with the environment.

Because decoherence acts, for macroscopic systems,
on an extremely short time scale, it appears to act discontinuously,
although in reality decoherence is a smooth process.
This is why ``events'', ``particles'', or ``quantum jumps'' are 
observed. Only in the special arrangement of experiments,
where systems are used that lie at the border between microscopic
and macroscopic, can this smooth nature of decoherence be observed.

Since decoherence studies only employ the standard formalism of
quantum theory, all components characterising macroscopically
different situations are still present in the total quantum state
which includes system {\em and} environment, although they cannot be 
observed locally. Whether there is a
real dynamical ``collapse'' of the total
state into one definite component or not (which would lead to an
Everett interpretation)
 is at present an undecided question.
Since this may not experimentally be decided in the near future,
it has been declared a ``matter of taste'' (Zeh~1997).

The most important feature of decoherence besides its ubiquity
is its {\em irreversible} nature. Due to the interaction with the
environment, the quantum mechanical entanglement {\em increases}
with time. Therefore, the local entropy for subsystems increases, too,
since information residing in correlations is locally unobservable.
A natural prerequisite for any such irreversible behaviour,
most pronounced in the Second Law of thermodynamics, is a special
initial condition of very low entropy. Penrose has 
demonstrated convincingly
 that this is due to the extremely special nature of the
big bang. Can this peculiarity be explained in any satisfactory way?
Convincing arguments have been put forward that this can only be
achieved within a quantum theory of gravity (Zeh~1999).
This leads directly into the realm of quantum cosmology
which is the topic of the following sections.

\section{Quantum cosmology}
\subsection{Why spacetime cannot be classical}
Quantum cosmology is the application of quantum theory
to the Universe as a whole. Is such a theory possible or
even -- as I want to argue here -- needed for consistency?
In the first section I have stressed the importance of the
superposition principle and the ensuing quantum entanglement
with environmental degrees of freedom. Since the environment is
in general also coupled to another environment, this leads 
ultimately to the whole Universe as the only closed quantum system
in the strict sense. Therefore one must take quantum cosmology
seriously. Since gravity is the dominant interaction
on the largest scales, one faces the problem of quantising
the gravitational field. In the following I shall list
some arguments that can be put forward in support of
such a quantisation, cf. Kiefer (1999):
\begin{itemize}
\item {\em Singularity theorems of general relativity}:
Under very general conditions, the occurrence of a singularity,
and therefore the breakdown of the theory,
is unavoidable. 
 A more fundamental theory is therefore needed to
overcome these shortcomings, and the general expectation is that this
fundamental theory is a quantum theory of gravity.
\item {\em Initial conditions in cosmology}: This is related to the
       singularity theorems,
      since they predict the existence of a
      ``big bang'' where the known laws of physics break down.
      To fully understand the evolution of our Universe, its initial
      state must be amenable to a physical description.
\item {\em Unification}: Apart from general relativity, all known
      fundamental theories are {\em quantum} theories. It would thus seem
      awkward if gravity, which couples to all other fields,
      should remain the only classical entity in a fundamental 
      description. Moreover, it seems that classical fields cannot
      be coupled to quantum fields without leading to inconsistencies
      (Bohr-Rosenfeld type of analysis).
\item {\em Gravity as a regulator}: Many models indicate that the consistent
      inclusion of gravity in a quantum framework automatically
      eliminates the divergences that plague ordinary quantum 
      field theory.
\item {\em Problem of time}: In ordinary quantum theory, the presence
       of an external time parameter $t$ is crucial for the interpretation
       of the theory: ``Measurements'' take place at a certain time,
       matrix elements are evaluated at fixed times, and the norm 
       of the wave function is conserved {\em in} time.
       In general relativity,
       on the other hand, time
       as part of spacetime  is a dynamical quantity. Both concepts of time
       must therefore be modified at a fundamental level.
       This will be discussed in some detail in the next subsection.
\end{itemize}

The task of quantising gravity has not yet been accomplished,
but approaches exist within which sensible questions can be asked.
Two approaches are at the centre of current research:
Superstring theory (or M-theory) and canonical quantum gravity.
Superstring theory is much more ambitious and aims at a unification
of all interactions within a single quantum framework
(a recent overview is Sen~1998).
Canonical quantum gravity, on the other hand, attempts to 
construct a consistent, non-perturbative, quantum theory of the
gravitational field on its own. This is done through the application
of standard quantisation rules to the general theory of relativity.

The fundamental length scales that are connected with these theories
are the Planck length, $l_p=\sqrt{G\hbar/c^3}$, or the
string length, $l_s$. It is generally assumed that the string length
is somewhat larger than the Planck length. Although not fully
established in quantitative detail, canonical quantum gravity
should follow from superstring theory for scales $l\gg l_s>l_p$.
One argument for this derives directly from the kinematical nonlocality
of quantum theory: Quantum effects are not a priori restricted
to certain scales. For example, the rather large mass of a dust grain
cannot by itself be used as an argument for classicality. Rather,
the process of decoherence through the environment can explain
{\em why} quantum effects are negligible for this object, see the 
discussion in Sect.~2.2, in particular the quantitative
aspects as they manifest themselves in the Table.
Analogously, the smallness of $l_p$ or $l_s$ cannot by itself be
used to argue that quantum-gravitational effects are small. Rather, this
should be an emergent fact to be justified by decoherence
(see Sect.~4). Since for scales larger than $l_p$ or $l_s$
general relativity is an excellent approximation, it must be
clear that the canonical quantum theory must be an excellent
approximation, too. The canonical theory might or might not
exist on a full, non-perturbative level, but it should definitely
exist as an effective theory on large scales.
It seems therefore sufficient to base the following discussion
on canonical quantum gravity, although I want to emphasise that
the same conceptual issues arise in superstring theory.

Depending on the choice of the canonical variables,
the canonical theory can be subdivided into the following
approaches:
\begin{itemize}
\item {\em Quantum geometrodynamics}: This is the traditional
      approach that uses the three-dimensional metric as
      its configuration variable. 
\item {\em Quantum connection dynamics}: The configuration variable
      is a non-abelian connection that has many similarities
      to gauge theories.
\item {\em Quantum loop dynamics}: The configuration variable is
      the trace of a holonomy with respect to a loop, analogous
      to a Wilson loop.
\end{itemize}  
There exists a connection between the last two approaches,
whereas their connection to the first approach is less clear.
 For the above reason one should, however,
expect that a relation between all approaches exists
at least on a semiclassical level. Here, I shall restrict myself to quantum
geometrodynamics, since this seems to be the most
appropriate language for a discussion of the conceptual issues.
However, most of this discussion should find its pendant
in the other approaches, too. A thorough discussion
of these other approaches can be found in many
contributions to this volume, see also Ashtekar (1999).

\subsection{Problem of time}
``Quantisation'' is a set of heuristic recipes which allows one
to guess the structure of the quantum theory from the 
underlying classical theory.
In the canonical approach, the first step is to identify the
canonical variables, the configuration and momentum variables
of the classical theory. Their Poisson brackets are then
translated into quantum operators. As a well-known theorem
by Groenewald and van~Hove states, such a translation is not
possible for most of the other variables. 

Details of the canonical formalism for general relativity
can be found in Isham (1992), Kucha\v{r} (1992), and the
references therein, and I shall give here only a brief
introduction. For the definition of the canonical momenta,
a time coordinate has to be distinguished. This spoils the
explicit four-dimensional covariance of general relativity --
the theory is reformulated to give a formulation for the
dynamics of {\em three}-dimensional hypersurfaces. It is then not
surprising that the configuration variable is the {\em three}-dimensional
metric, $h_{ab}({\vec x})$, on such hypersurfaces. The three-metric
has six independent degrees of freedom. The remaining four
components of the spacetime metric play the role of non-dynamical
Lagrange multipliers called lapse function, $N^{\perp}({\vec x})$,
and shift vector, $N^a({\vec x})$ -- they parametrise, respectively, the
way in which consecutive hypersurfaces are chosen and
how the coordinates are selected {\em on} a hypersurface.
The momenta canonically conjugated to the three-metric,
$p^{ab}({\vec x})$,
form a tensor which is linearly related to the second fundamental form
associated with a hypersurface -- specifying the way in which
the hypersurface is embedded into the fourth dimension.
In the quantum theory, the canonical variables are formally
turned into operators obeying the commutation relations
\be
[\hat{h}_{ab}({\vec x}),\hat{p}^{ab}({\vec y})]
=\I\hbar\delta^c_{(a}\delta^d_{b)}\delta({\vec x},{\vec y})\ .
\ee
In a (formal) functional Schr\"odinger representation, the
canonical operators act on wave functionals $\Psi$ depending
on the three-metric,
\bea
\hat{h}_{ab}({\vec x})  \Psi[h_{ab}({\vec x})]&=&
 h_{ab}({\vec x})  \Psi[h_{ab}({\vec x})]\\
\hat{p}^{cd}({\vec x}) \Psi[h_{ab}({\vec x})] &=&
\frac{\hbar}{\I}\frac{\delta}{\delta h_{cd}({\vec x})}
 \Psi[h_{ab}({\vec x})]\ .
\eea
A central feature of canonical gravity is the existence of
constraints. Because of the four-dimensional diffeomorphism
invariance of general relativity, these are four constraints
per space point, one Hamiltonian constraint,
\be 
\hat{\cal H}_{\perp}\Psi=0\ , \lb{ham}
\ee
and three diffeomorphism constraints,
\be
\hat{\cal H}_a\Psi=0\ . \lb{diffeo}
\ee
The total Hamiltonian is obtained by integration\footnote{In
the following I shall restrict myself to closed compact spaces;
otherwise, the Hamiltonian has to be augmented by surface terms
such as the ADM energy.},
\be
\hat{H}=\int\D^3x\ (N^{\perp}\hat{\cal H}_{\perp}
         +N^a\hat{\cal H}_a), \ \lb{totham}
\ee
where $N^{\perp}$ and $N^a$ denote again lapse function and
shift vector, respectively. The constraints then enforce that
the wave functional be annihilated by the total Hamiltonian,
\be
\hat{H}\Psi=0\ . \lb{wdw}
\ee
The {\em Wheeler-DeWitt} equation (\ref{wdw}) is the central equation
of canonical quantum gravity. This also holds for quantum connection
dynamics and quantum loop dynamics, although the configuration
variables are different. 

The Wheeler-DeWitt equation (\ref{wdw}) possesses the remarkable
property that it does not depend on any external time parameter --
the $t$ of the time-dependent Schr\"odinger equation has totally
disappeared, and (\ref{wdw}) looks like a stationary zero-energy
Schr\"odinger equation. How can this be understood?
In classical canonical gravity, a spacetime can be represented  
as a ``trajectory'' in configuration space -- the space of all
three-metrics. Although time coordinates have no intrinsic meaning
in classical general relativity either, they can nevertheless
be used to parametrise this trajectory in an essentially arbitrary
way. Since no trajectories exist anymore in quantum theory,
no spacetime exists at the most fundamental, and therefore also
no time coordinates to parametrise any trajectory. 
A simple analogy is provided by the relativistic particle:
In the classical theory there is a trajectory which can be
parametrised by some essentially arbitrary parameter, e.g. 
the proper time. Reparametrisation invariance
leads to one constraint, $p^2+m^2=0$. In the quantum theory,
no trajectory exists anymore, the wave function obeys 
the Klein-Gordon equation as an analogue of (\ref{wdw}),
and any trace of a classical time parameter is lost
(although, of course, for the relativistic particle the 
background Minkowski spacetime is present, which is
not the case for gravity).

Since the presence of an external time parameter is very 
important in quantum mechanics -- giving rise to such important
notions as the unitarity of states --, it is a priori not clear
how to interpret a ``timeless'' equation of the form (\ref{wdw}),
cf. Barbour (1997) and Kiefer (1997). This is called the {\em problem of time}.
A related issue is the {\em Hilbert-space problem}: What is the appropriate
inner product that encodes the probability interpretation
and that is conserved in time?
Before discussing some of the options, it is very useful to first
have a look at the explicit structure of (\ref{ham}) and
(\ref{diffeo}). Introducing the Planck mass $m_p=(16\pi G)^{-1/2}$
and setting $\hbar=1$, the constraint equations read
        \begin{eqnarray}
        &&\mbox{\vphantom{$\left\{\frac{L^{L^{L^a}}_{L}}
        {L^L_L g}\right\}$}}^{\prime\prime}\!
        \left\{-\frac 1{2 m_p^2}
        G_{ab,cd}\frac{\delta^2}
        {\delta h_{ab}\delta h_{cd}}-m_p^2\,\sqrt{h}\,{}^3\!R +
        \hat{H}{}_{\perp}^{\rm mat}\right\}^{\prime\prime}
        |{\mbox{\boldmath$\Psi$}}[h_{ab}]\big>=0,           \label{ham1}
        \\
        &&\mbox{\vphantom{$\left\{\frac{L^{L^{L^a}}_{L}}
        {L^L_L g}\right\}$}}^{\prime\prime}\!
        \left\{-\frac{2}\I h_{ab}\nabla_c
        \frac{\delta}{\delta h_{bc}}+\hat{H}{}_a^{\rm mat}
        \right\}^{\prime\prime}
        |{\mbox{\boldmath$\Psi$}}[h_{ab}]\big>=0.            \label{diffeo1}
        \end{eqnarray}  
 The inverted commas indicate that these are formal equations and that
the factor ordering and regularisation problem have not been
addressed. In these equations, $^3\!R$ and $\sqrt{h}$ denote the
three-dimensional Ricci scalar and the square root of the
determinant of the
three-metric, respectively, and a cosmological term has not been
considered here. The quantity $G_{ab,cd}=h^{-1/2}
(h_{ac}h_{bd}+h_{ad}h_{bc}-h_{ab}h_{cd})$ plays the role of
a metric in configuration space (``DeWitt metric''), and $\nabla_c$
denotes the covariant spatial derivative.
 The matter parts of the constraints, $\hat{H}_{\perp}^{\rm mat}$ and
$\hat{H}_a^{\rm mat}$, depend on the concrete choice of matter action
which we shall not specify here. Its form can be strongly constrained
from general principles such as ultralocality (Teitelboim~1980).
A tilde denotes a quantum operator in the standard Hilbert space
of matter fields, while the bra and ket notation refers to
the corresponding states. 

The second equation (\ref{diffeo1}) expresses the fact that
the wave functional is invariant with respect to three-dimensional
diffeomorphisms (``coordinate transformations''). It is
for this reason why one often writes $\Psi[^3\!{\cal G}]$,
where the argument denotes the coordinate-invariant
three-{\em geometry}. Since there is, however, no explicit
operator available which acts directly on $\Psi[^3\!{\cal G}]$,
this is only a formal representation, and in concrete
discussions one has to work with (\ref{ham1}) and
(\ref{diffeo1}). It must also be remarked that this invariance
holds only for diffeomorphisms that are connected with the
identity; for ``large'' diffeomeorphism, a so-called
$\theta$-structure may arise, similarly to the $\theta$-angle
in QCD, see e.g. Kiefer (1993). 

The kinetic term in (\ref{ham1}) exhibits an interesting
structure: The DeWitt metric $G_{ab,cd}$ has locally
the signature $\mbox{diag}(-,+,+,+,+,+)$, rendering the kinetic
term {\em indefinite}. Moreover, the one {\em minus sign} in
the signature suggests that the corresponding degree
of freedom plays the role of an ``intrinsic time''
(Zeh~1999). In general this does not, however, 
render (\ref{ham1}) a hyperbolic equation, since even after 
dividing out the diffeomorphisms -- going to
the {\em superspace} of all three-geometries -- there
remains in general an infinite number of minus signs.
In the special, but interesting, case of perturbations around
closed Friedmann cosmologies, however, one global
minus sign remains, and one is left with a truly
hyperbolic equation (Giulini~1995). A Cauchy problem
with respect to intrinsic time may then be posed.
The minus sign in the DeWitt metric can be associated with
the local scale part, $\sqrt{h}$, of the three-metric.

The presence of the minus sign in the DeWitt metric has
an interesting interpretation: It reflects the fact
that gravity is {\em attractive} (Giulini and Kiefer~1994).
This can be investigated by considering the most general
class of ultralocal DeWitt metrics which are characterised
by the occurrence of some additional parameter $\alpha$:
\be
G_{ab,cd}^{\alpha}=h^{-1/2}(h_{ac}h_{bd}+
h_{ad}h_{bc}-2\alpha h_{ab}h_{cd})\ ,
\ee 
where $\alpha=0.5$ is the value corresponding to general relativity.
One finds that there exists a critical value,
$\alpha_c=1/3$, such that for $\alpha<\alpha_c$ the DeWitt metric
would become positive definite. One also finds that for
$\alpha<\alpha_c$ gravity would become repulsive in the
following sense: First, the second time derivative of the
total volume $V=\int \D^3x\sqrt{h}$ (for lapse equal to one)
would become, for positive three-curvature, positive instead of
negative, therefore leading to an acceleration. Second,
in the coupling to matter the sign of the gravitational constant
would change. From the observed amount of helium one can
infer that $\alpha$ must lie between 0.4 and 0.55. 

Standard quantum theory employs the mathematical structure
of a Hilbert space which is needed for the probability
interpretation. Does such a structure also exist in
quantum gravity? On a kinematical level, for wave functionals
which are not yet necessarily solutions of the constraint
equations, one can try to start with the standard
Schr\"odinger-type inner product
\be
\int{\cal D}h_{ab}\Psi^*[h_{ab}({\vec x})]
     \Psi[h_{ab}({\vec x})]\equiv (\Psi,\Psi)_S\ . \lb{sp}
\ee
For wave functionals which satisfy the diffeomorphism constraints
(\ref{diffeo1}), this would yield divergencies since the
integration runs over all ``gauge orbits''. In the connection
representation, a preferred measure exists with respect to which
the wave functionals are square integrable functions on the
space of connections, see the contributions by Ashtekar,
Lewandowski, and Rovelli to this volume. The construction
is possible because the Hilbert space can be viewed as a limit
of Hilbert spaces with finitely many degrees of freedom.
It leads to interesting results for the spectra of geometric
operators such as the area operator. However, no such
product is known in geometrodynamics.

Since physical wave functionals have to obey (\ref{ham1})
{\em and} (\ref{diffeo1}), it might be sufficient if a Hilbert-space
structure existed on the space of solutions, not necessarily
on the space of all functionals such as in (\ref{sp}).
Since (\ref{ham1}) has locally the form of a Klein-Gordon equation,
one might expect to use the inner product
\be
\I\int\Pi_{\vec x}\D\Sigma^{ab}({\vec x})\Psi^*[h_{ab}]
 \left(G_{ab,cd}\stackrel{\rightarrow}{\frac{\delta}
  {\delta h_{cd}}}-\stackrel{\leftarrow}{\frac{\delta}
  {\delta h_{cd}}}G_{ab,cd}\right)\Psi[h_{ab}]\equiv
  (\Psi,\Psi)_{KG}\ . \lb{kg}
\ee
The (formal) integration runs over a five-dimensional 
hypersurface at each space point, which is spacelike
with respect to the DeWitt metric. The product (\ref{kg})
is invariant with respect to deformations of this 
hypersurface and therefore independent of ``intrinsic time''.

Similar to the situation with the relativistic particle,
however, the inner product (\ref{kg}) is {\em not}
positive definite. For the {{\em free} relativistic particle 
one can perform a consistent restriction to a 
``positive-frequency sector'' in which the analogue of (\ref{kg})
is manifestly positive, {\em provided} the spacetime background
and the potential (which must be positive) are stationary, i.e.,
if there exists a time-like Killing vector which also preserves
the potential. Otherwise, ``particle production'' occurs
and the one-particle interpretation of the theory cannot be maintained.
It has been shown that such a restriction to ``positive
frequencies'' is {\em not} possible in quantum
geometrodynamics (Kucha\v{r}~1992), the reason being that
the Hamiltonian is not stationary. As I shall describe in Sect.~4,
one can make, at least for certain states in the ``one-loop level'' 
of the semiclassical approximation, a consistent
restriction to a positive-definite sector of (\ref{kg}).

For the relativistic particle one leaves the one-particle
sector and proceeds to a field-theoretic setting, if
one has to address situations where the restriction to
positive frequencies is no longer possible. One then arrives
at wave {\em functionals} for which a Schr\"odinger-type of inner
product can be formulated.
 Can one apply
a similar procedure for the Wheeler-DeWitt equation?
Since quantum geometrodynamics is already a field theory,
this would mean performing the transition to a ``third-quantised''
theory in which the state in (\ref{wdw}) is itself turned into
an operator. The formalism for such a theory is still in
its infancy and will not be presented here
 (see e.g. Kucha\v{r}~1992). In a sense, superstring
theory can be interpreted as providing such a framework.

All these problems could be avoided if it were possible
to ``solve'' the constraints classically and make a
transition to the physical degrees of freedom,
upon which the standard Schr\"odinger inner product
could be imposed. This would
correspond to the choice of a time variable before quantisation.
Formally, one would have to perform the canonical transformation
\be
(h_{ab},p^{cd}) \longrightarrow (X^A,P_A;\phi^i,p_i)\ , \lb{can}
\ee
where $A$ runs from 1 to 4, and $i$ runs from 1 to 2.
$X^A$ and $P^A$ are the kinematical ``embedding variables'',
while $\phi^i$ and $p_i$ are the dynamical, physical, degrees
of freedom. Unfortunately, such a reduction can only be performed
in special situations, such as weak gravitational waves, but not
in the general case, see Isham (1992) and Kucha\v{r} (1992). 
The best one can do is to choose the so-called
``York time'', but the corresponding reduction cannot be
performed explicitly. Again, only on the one-loop level
of the semiclassical approximation (see Sect.~4) can the
equivalence of the Schr\"odinger product for the reduced
variables and the Klein-Gordon inner product for the constrained
variables be shown. 

The problems of time and Hilbert space are thus not yet resolved
at the most fundamental level. It is thus not clear, for example,
whether (\ref{wdw}) can sensibly be interpreted only as an eigenvalue
equation for eigenvalue zero.
Thus the options that will be discussed
in the rest of my contribution are  
\begin{itemize}
\item to study a semiclassical approximation and to aim at
      a consistent treatment of conceptual issues at that level.
      This is done in Sect.~4. Or
\item to look for sensible boundary conditions for the
      Wheeler-DeWitt equation and to discuss directly solutions
      to this equation. This is done in the rest of this section.
\end{itemize}

\subsection{Role of boundary conditions}
Boundary conditions play a different role in quantum mechanics
and quantum cosmology. In quantum mechanics (more generally,
quantum field theory with an external background), boundary
conditions can be imposed with respect to the external
time parameter: Either as a condition on the wave function
at a given time, or as a condition on asymptotic states
in scattering situations. On the other hand, the
Wheeler-DeWitt equation (\ref{wdw}) is a ``timeless'' equation
with a Klein-Gordon type of kinetic term. 

What is the role of boundary conditions in quantum cosmology?
Since the time of Newton one is accustomed to distinguish
between dynamical laws and initial conditions. However, 
this is not a priori clear in quantum cosmology, and it might
well be that boundary conditions are part of the
dynamics. Sometimes quantum cosmology is even called
a {\em theory of initial conditions} (Hartle~1997). Certainly, ``initial''
can here have two meanings: On the one hand, it can refer
to initial condition of the classical Universe. This
presupposes the validity of a semiclassical approximation
(see Sect.~4) and envisages that particular solutions
of (\ref{wdw}) could {\em select} a subclass of classical solutions
in the semiclassical limit. On the other hand, ``initial''
can refer to boundary conditions being imposed directly
on (\ref{wdw}). Since (\ref{wdw}) is fundamentally timeless,
 this cannot refer
to any classical time parameter but only to intrinsic variables
such as ``intrinsic time''. In the following I shall briefly
review some boundary conditions that have been suggested
in quantum cosmology; details and additional references
can be found in Halliwell (1991). 

Let me start with the no-boundary proposal by Hartle and
Hawking (1983). This does not yield directly boundary conditions
on the Wheeler-DeWitt equation, but specifies the wave function
through an integral expression -- through a path integral
in which only a subclass of all possible ``paths'' is
being considered. This subclass comprises all spacetimes that
have (besides the boundary where the arguments of the wave
function are specified) no other boundary. Since the full
quantum-gravitational path integral cannot be evaluated
(probably not even be rigorously defined), one must resort
to approximations. These can be semiclassical or minisuperspace
approximations or a combination of both. It becomes clear
already in a minisuperspace approximation that integration
has to be performed over {\em complex} metrics to guarantee
convergence. Depending on the nature of the saddle point
in a semiclassical limit, the wave function can then refer to
a classically allowed or forbidden situation. 

Consider the example of a Friedmann Universe with a 
conformally coupled scalar field. After an appropriate
field redefinition, the Wheeler-DeWitt equation assumes
the form of an indefinite harmonic oscillator,
\be
\left(\frac{\partial^2}{\partial a^2}- \frac{\partial^2}
          {\partial\phi^2}-a^2+\phi^2\right)
          \psi(a,\phi)=0 \ .
\ee
The implementation of the no-boundary condition in this
simple minisuperspace model selects the following solutions
(cf. Kiefer~1991)
\bea
\psi_1(a,\phi) &=& \frac{1}{2\pi}K_0\left(\frac{|\phi^2-a^2|}{2}
      \right)\ , \\
\psi_2(a,\phi) &=&  \frac{1}{2\pi}I_0\left(\frac{\phi^2-a^2}{2}
       \right)\ ,
\eea
where $K_0$ and $I_0$ denote Bessel functions.
It is interesting to note that these solutions do not
reflect the classical behaviour of the system (the classical
solutions are Lissajous ellipses confined to a rectangle in
configuration space, see Kiefer~1990) -- $I_0$ diverges for
large arguments, while $K_0$ diverges for vanishing argument
(``light cone'' in configuration space). Such features cannot
always be seen in a semiclassical limit. 

Another boundary condition is the so-called
tunneling condition (Vilenkin 1998). It is also formulated
in general terms -- superspace should contain ``outgoing
modes'' only. However, as with the no-boundary proposal, 
a concrete discussion can only be made within approximations.
Typically, while the no-boundary proposal leads to {\em real}
solutions of the Wheeler-DeWitt equation, the tunneling proposal
predicts {\em complex} solutions. This is most easily seen in
the semiclassical approximation (see Sect.~4), where the former
predicts $\cos S$-type of solutions, while the latter predicts
$\exp\I S$-type of solutions. (The name ``tunneling proposal''
comes from the analogy with situations such as $\alpha$-decay
in nuclear physics where an outgoing wave is present
after tunneling from the nucleus.) A certain danger is connected
with the word ``outgoing'' because it has a temporal connotation
although (\ref{wdw}) is timeless. A time parameter emerges only in 
a semiclassical approximation, see the next section.

A different type of boundary condition is the SIC proposal
by Conradi and Zeh (1991). It demands that the wave function
be simple for small scale factors, i.e. that it does not
depend on other degrees of freedom. The explicit expressions
exhibit many similarities to the no-boundary wave function,
but since the boundary condition is directly imposed
on the wave function without use of path integrals,
it is much more convenient for a discussion of models
which correspond to a classically recollapsing universe. 

What are the physical applications that one could
possibly use to distinguish between the various boundary
conditions? Some issues are the following:
\begin{itemize}
\item {\em Probability for inflation}: It is often assumed that the
 Universe underwent a period of exponential expansion at
 an early stage (see also Sect.~4.3). The question therefore
  arises whether quantum cosmology can predict how ``likely''
 the occurrence of inflation is. Concrete calculations
 address the question of the probability distribution for
 the initial values of certain fields that are responsible
 for inflation. Since such calculations necessarily involve
 the validity of a semiclassical approximation (otherwise
 the notion of inflation would not make sense), I shall 
 give some more details in the next section.
\item {\em Primordial black-hole production}: The production of primordial
 black holes during an inflationary period can in principle
 also be used to discriminate between boundary conditions, see
 e.g. Bousso and Hawking (1996). 
\item {\em Cosmological parameters}: If the wave function is peaked
 around definite values of fundamental fields, these values
 may appear as ``constants of Nature'' whose values can thereby be
 predicted. This was tentatively done for the
cosmological constant (Coleman~1988). Alternatively, the
anthropic principle may be invoked to select amongst the
values allowed by the wave function.
\item {\em Arrow of time}: Definite conclusions about the
arrow of time in the Universe (and the interior of black
holes) can be drawn from solutions to the Wheeler-DeWitt
equation, see Kiefer and Zeh (1995).
\end{itemize}
Quantum cosmology is of course not restricted to quantum
general relativity. It may also be discussed within
effective models of string theory, see e.g. D\c{a}browski
and Kiefer (1997), but I shall not discuss this here.
\section{Emergence of a classical world}
As I have reviewed in Sect.~3, there is no notion of
spacetime at the full level of quantum cosmology. This was
aleady anticipated by Lema\^{\i}tre (1931) who wrote:
\begin{quote}
If the world has begun with a single quantum, the notions
of space and time would altogether fail to have any meaning
at the beginning \ldots If this suggestion is correct,
the beginning of the world happened a little before
the beginning of space and time.
\end{quote}
It is not clear what ``before'' means in an atemporal
situation, but it is obvious that the emergence of the
usual notion of spacetime within quantum cosmology needs
an explanation. This is done in two steps: Firstly,
a semiclassical approximation to quantum gravity must
be performed (Sect.~4.1). This leads to the recovery
of an {\em approximate} Schr\"odinger equation of non-gravitational
fields with respect to the semiclassical background.
Secondly, the emergence of classical properties
must be explained (Sect.~4.2). This is achieved through the
application of the ideas presented in Sect.~2.2.
A more technical review is Kiefer (1994),
see also Brout and Parentani (1999).
A final subsection is devoted to the emergence of classical
fluctuations which can serve as seeds for the
origin of structure in the Universe.

\subsection{Semiclassical approximation to quantum gravity}
The starting point is the observation that there occur
different scales in the fundamental equations (\ref{ham1})
and (\ref{diffeo1}):
The Planck mass $m_p$ associated with the gravitational part,
and other scales contained implicitly in 
$\hat{H}{}^{\rm mat}_{\perp}$. Even for ``grand-unified
theories'' the relevant particle scales are at least
three orders of magnitude smaller than $m_p$. For this reason
one can apply Born-Oppenheimer type of techniques that are
suited to the presence of different scales. In molecular
physics, the large difference between nuclear mass and  
electron mass leads to a slow motion for the nuclei
and the applicability of an adiabatic approximation.
A similar method is also applied in the nonrelativistic
approximation to the Klein-Gordon equation,
see Kiefer and Singh (1991).

In the lowest order of the semiclassical approximation,
the wave functional appearing in (\ref{ham1}) and (\ref{diffeo1})
can be written in the form
        \begin{eqnarray}
        |{\mbox{\boldmath$\Psi$}}[h_{ab}]\big>=
        \E^{\,{}^{\textstyle{\I m_p^2
        {\mbox{\boldmath$S$}[h_{ab}]}}}}|\Phi [h_{ab}]\big>\ , \label{semi}
        \end{eqnarray}
where ${\mbox{\boldmath$S$}[h_{ab}]}$ is a purely gravitational
Hamilton-Jacobi function. This is a solution of the vacuum
Einstein-Hamilton-Jacobi equations -- the gravitational constraints
with the Hamilton-Jacobi values of momenta (gradients
of ${\mbox{\boldmath$S$}[h_{ab}]}$).

Substitution of (\ref{semi}) into (\ref{ham1}) and (\ref{diffeo1}) leads
to new equations for the state vector of matter fields $|\Phi[h_{ab}]\big>$
depending parametrically on the spatial metric
        \begin{eqnarray}
        &&\left \{\frac 1\I G_{ab,cd}
        \frac{\delta{\mbox{\boldmath$S$}}}
        {\delta h_{ab}} \frac{\delta}
        {\delta h_{cd}}+\hat{H}_{\perp}^{\rm mat}(h_{ab})\right.\nonumber\\
        &&\, 
        \left.+\frac 1{2\I}\,
        \mbox{\vphantom{$\left\{\frac{L^{L^{L^a}}_{L}}
        {L^L_L g}\right\}$}}^{\prime\prime}\!\!
        G_{ab,cd}
        \frac{\delta^2{\mbox{\boldmath$S$}}}
        {\delta h_{ab}\delta h_{cd}}^{\prime\prime}
        -\frac 1{2m_p^2}\,\mbox{\vphantom{$\left\{\frac{L^{L^{L^a}}_{L}}
        {L^L_L g}\right\}$}}^{\prime\prime}\!\!
        G_{ab,cd}\frac{\delta^2}
        {\delta h_{ab}\delta h_{cd}}^{\prime\prime}
        \right\}
        |\Phi[h_{ab}]\big>=0,                  \label{hsemi}\\
        &&\left\{\mbox{\vphantom{$\left\{\frac{L^{L^{L^a}}_{L}}
        {L^L_L g}\right\}$}}^{\prime\prime}\!\!
        -\frac{2}\I h_{ab}\nabla_c
        \frac{\delta}{\delta h_{bc}}^{\prime\prime}+
        \hat{H}_a^{\rm mat}(h_{ab})\right\}|\Phi[h_{ab}]\big>=0. \label{dsemi}
        \end{eqnarray}

It should be emphasised that on a formal level the factor
ordering can be {\em fixed} by demanding the equivalence
of various quantisation schemes, see Al'tshuler and Barvinsky (1996)
and the references therein.
 
The conventional derivation of the Schr\"odinger equation from the
Wheeler-DeWitt equation consists in the assumption of {\it small
back reaction of quantum matter on the metric background} which
at least heuristically allows one to discard the third and the fourth
terms in (\ref{hsemi}). Then one considers $|\Phi[h_{ab}]\big>$ on the
solution of classical vacuum Einstein equations $h_{ab}({\bf x},t)$
corresponding to the Hamilton-Jacobi function
 ${\mbox{\boldmath$S$}[h_{ab}]}$,
        $|\Phi(t)\big>=|\Phi[h_{ab}({\bf x},t)]\big>$.
After a certain {\em choice} of lapse and shift functions $(N^{\perp},N^a)$,
this solution satisfies the canonical equations with the momentum
$p^{ab}=\delta{\mbox{\boldmath$S$}}/\delta h_{ab}$, so that the quantum
state $|\Phi(t)\big>$ satisfies the evolutionary equation obtained by using
        \begin{eqnarray}
        \frac{\partial}{\partial t}\,|\Phi(t)\big>=
        \int \D^3 x \,\dot{h}_{ab}({\bf x})\,
        \frac{\delta}{\delta h_{ab}({\bf x})}
        |\Phi[h_{ab}]\big>
        \end{eqnarray}
together with the truncated version of equations (\ref{hsemi})
 -- (\ref{dsemi}).
The result is the Schr\"odinger equation of quantised matter fields in the
external classical gravitational field,
        \begin{eqnarray}
        &&\I\frac{\partial}{\partial t}\,
        |\Phi(t)\big>=\hat{H}{}^{\rm mat}|\Phi(t)\big>,  \label{sch}\\
        &&\hat{H}{}^{\rm mat}=
        \int \D^3 x \left\{N^{\perp}({\bf x})
        \hat{H}{}^{\rm mat}_{\perp}({\bf x})+
        N^a({\bf x})\hat{H}{}^{\rm mat}_a({\bf x})\right\}.     \label{Hmat}
        \end{eqnarray}
Here, $\hat{H}{}^{\rm mat}$ is a matter field Hamiltonian in the Schr\"odinger
picture, parametrically depending on (generally nonstatic) metric
coefficients of the curved spacetime background. In this way,
the Schr\"odinger equation for non-gravi\-ta\-tio\-nal fields
has been recovered from quantum gravity as an approximation.

A derivation similar to the above can already be performed
within ordinary quantum mechanics if one assumes that the
total system is in a ``timeless'' energy eigenstate,
see Briggs and Rost (1999). In fact, Mott (1931) had already
considered a time-independent Schr\"odinger equation
for a total system consisting of an $\alpha$-particle
and an atom. If the state of the $\alpha$-particle can be 
described by a plane wave (corresponding in this case to
high velocities), one can make an ansatz similar to (\ref{semi})
and derive a time-dependent Schr\"odinger equation for the
atom alone, in which time is {\em defined} by the $\alpha$-particle.

In the context of quantum gravity, it is most
interesting to continue the semiclassical approximation
to higher orders and to {\em derive} quantum-gravitational
correction terms to (\ref{sch}). This was done in
Kiefer and Singh (1991) and, giving a detailed interpretation
in terms of a Feynman diagrammatic language, in
Barvinsky and Kiefer (1998). I shall give a brief description
of these terms and refer
the reader to Barvinsky and Kiefer (1998) for all details.

At the next order of the semiclassical expansion, one obtains
corrections to (\ref{sch}) which are proportional
to $m_p^{-2}$. These terms can be added to the matter Hamiltonian,
leading to an {\em effective} matter Hamiltonian at this order.
It describes the back-reaction effects of quantum matter on
the dynamical gravitational background as well as proper
quantum effects of the gravitational field itself.
Most of these terms are nonlocal in character:
they contain the gravitational potential generated by the
back reaction of quantum matter as well as the gravitational
potential generated by the one-loop stress tensor of
vacuum gravitons. In cases where the matter energy density
is much bigger than the energy density of graviton vacuum
polarisation, the dominant correction term is given by the
kinetic energy of the gravitational radiation produced
by the back reaction of quantum matter sources.

A possible observational test of these correction terms
could be provided by the anisotropies in the cosmic
microwave background (Rosales~1997). The temperature fluctuations
are of the order $10^{-5}$ reflecting within inflationary models
the ratio $m_I/m_p\approx 10^{-5}$, where $m_I$ denotes 
the mass of the scalar field responsible for inflation
(the ``inflaton'').
The correction terms would then be $(m_I/m_p)^2\approx 10^{-10}$
times a numerical constant, which could in principle be large enough
to be measurable with future satellite experiments such as
MAP or PLANCK.

Returning to the ``one-loop order'' (\ref{semi}) of the
semiclassical approximation, it is possible to address the
issue of probability for inflation that was mentioned in
Sect.~3.3, see Barvinsky and Kamenshchik (1994).
 In this approximation, the inner products
(\ref{sp}) and (\ref{kg}) are equivalent and positive
definite, see Al'tshuler and Barvinsky (1996). They can
therefore be used to calculate quantum-mechanical
probabilities in the usual sense.

To discuss this probability, the reduced density matrix for
the inflaton,
$\varphi$, should be investigated. This density matrix
is calculated from the full quantum state upon integrating out 
all other degrees of freedom (here called $f$),
\be \rho_t(\varphi,\varphi')=\int{\cal D}f\
       \psi_t^*(\varphi',f)\psi_t(\varphi,f)\ , \lb{rho} \ee
where $\psi_t$ denotes the quantum state (\ref{semi})
after the parameter $t$ from (\ref{sch}) has been used.

To calculate the probability one has to set $\varphi'=\varphi$.
In earlier work, the saddle-point approximation was only performed
up to the highest, tree-level, approximation.
This yields
\be \rho(\varphi,\varphi)=\exp[\pm I(\varphi)]\ , \lb{tree} \ee
where $I(\varphi)=-3m_p^4/8V(\varphi)$ and
$V(\varphi)$ is the inflationary poential. The lower sign corresponds
to the no-boundary condition, while the upper sign corresponds to
the
tunneling condition. The problem with (3) is that $\rho$ is not
normalisable: mass scales bigger than $m_p$ contribute significantly
and results
based on tree-level approximations can thus not be trusted.

The situation is improved considerably if
loop effects are taken into account (Barvinsky and Kamenshchik~1994).
They are incorporated by the loop effective action $\Gamma_{loop}$
which is calculated on 
De-Sitter space. In the limit of large $\varphi$
(that is relevant for investigating normalisability)
this yields in the one-loop approximation
\be \Gamma_{loop}(\varphi)\vert_{H\to\infty}
        \approx Z\ln\frac{H}{\mu}\ , \ee
where $\mu$ is a renormalisation mass parameter, and $Z$ is the
anomalous scaling.
Instead of (\ref{tree}) one has now 
\bea \rho(\varphi,\varphi)&\approx& \ H^{-2}(\varphi)
     \exp\left(\pm I(\varphi)-\Gamma_{loop}(\varphi)\right)
     \nonumber\\
    &\approx& \ \exp\left(\pm\frac{3m_p^4}{8V(\varphi)}\right)
    \varphi^{-Z-2} \ . \eea
This density matrix is normalisable provided $Z>-1$.
This in turn leads to reasonable constraints on the
particle content of the theory, see Barvinsky and Kamenshchik (1994).
It turns out that the tunneling wave function (with an
appropriate particle content) can predict the occurrence
of a sufficient amount of inflation.
In earlier tree-level calculations the use of an anthropic principle
was needed to get a sensible result from a non-normalisable wave
function through {\em conditional} probabilities, see
e.g. Hawking and Turok (1998). This is no longer the case here.

\subsection{Decoherence in quantum cosmology\protect
    \footnote{This
 and the next subsection are adapted from Kiefer (1999).}}
As in ordinary quantum mechanics, the semiclassical limit
is not yet sufficient to understand classical behaviour.
Since the superposition principle is also valid in 
quantum gravity, quantum entanglement will easily occur,
leading to superpositions of ``different spacetimes''. 
It is for this reason that the process of decoherence
must be invoked to justify the emergence of a classical
spacetime.

Joos (1986) gave a heuristic
example within Newtonian (quantum) gravity, in which the superposition
of different metrics is suppressed by the interaction with
ordinary particles. How does decoherence work in quantum cosmology?
In particular, what constitutes system and environment in a case
where nothing is external to the Universe? The question is
how to divide the degrees of freedom in the configuration
space in a sensible way. It was suggested by Zeh (1986)
to treat global degrees of freedom such as the scale factor
(radius) of the Universe or an inflaton field as ``relevant''
variables that are decohered by ``irrelevant'' variables
such as density fluctuations, gravitational waves, or other
fields. Quantitative calculations can be found, e.g., in
Kiefer (1987,1992). 

Denoting the ``environmental'' variables collectively again by
$f$, the reduced density matrix for e.g. the scale factor $a$
is found in the usual way by integrating out the $f$-variables,
\be
\rho(a,a')=\int{\cal D}f\ \Psi^*(a',f)\Psi(a,f)\ . \lb{nondiag}
\ee
In contrast to the discussion following (\ref{rho}),
the {\em non-diagonal} elements of the density matrix must be
calculated.
The resulting terms are ultraviolet-divergent and must therefore
be regularised. This was investigated in detail for the
case of bosons (Barvinsky et al.~1999c) and fermions
(Barvinsky et al.~1999a). A crucial point is that standard
regularisation schemes, such as dimensional regularisation
or $\zeta$-regularisation, do not work -- they lead to
$\mbox{Tr}\rho^2=\infty$, since the sign in the
exponent of the Gaussian density matrix is changed
from minus to plus by regularisation. 
These schemes therefore spoil one of the
important properties that a density matrix must obey. 
This kind of problem has not been noticed before, since
these regularisation schemes had not been applied 
to the calculation of reduced density matrices.

How, then, can (\ref{nondiag}) be regularised? In Barvinsky et al. (1999a,c)
we put forward the principle that there should be no decoherence
if there is no particle creation  -- decoherence is an
irreversible process. In particular, there should be no decoherence
for static spacetimes. This has led to the use of a certain
conformal reparametrisation for bosonic fields and
a certain Bogoliubov transformation for fermionic fields.

As a concrete example, we have calculated the reduced density matrix
for a situation where the semiclassical background
is a De~Sitter spacetime, $a(t)=H^{-1}\cosh(Ht)$,
where $H$ denotes the Hubble parameter. This is the 
most interesting example for the early Universe, since it is
generally assumed that there happened such an exponential,
``inflationary'', phase of the Universe, caused by an effective
cosmological constant. Taking various ``environments'', the following
results are found for the main contribution to
(the absolute value of) the decoherence
factor, $|D|$, that multiplies the reduced density matrix for
the ``isolated'' case:
\begin{itemize}
\item {\em Massless conformally-invariant field}: Here,
\[ |D|=1\ , \]
since no particle creation and therefore no decoherence effect
takes place. 
\item {\em Massive scalar field}: Here,
\[ |D|\approx\exp\left(-\frac{\pi m^3a}{128}(a-a')^2\right)\ , \]
and one notices increasing decoherence for increasing $a$.
\item {\em Gravitons}: This is similar to the previous case, but
the mass $m$ is replaced by the Hubble parameter $H$,
\[ |D|\approx\exp\left(-CH^3a(a-a')^2\right)\ , \; C>0\ . \]
\item {\em Fermions}:
\[ |D|\approx\exp\left(-C'm^2a^2H^2(a-a')^2\right)\ , \; C'>0\ . \]
For high-enough mass, the decoherence effect by fermions
is thus smaller than the corresponding influence of bosons.
\end{itemize}
It becomes clear from these examples that the Universe
acquires classical properties after the onset of the
inflationary phase. ``Before'' this phase, the Universe
was in a timeless quantum state which does not possess any
classical properties. Viewed backwards, different semiclassical
branches would meet and interfere to form this timeless
quantum state (Barvinsky et al.~1999b). 

For these considerations it is of importance that 
there is a discrimination between the various degrees of freedom.
On the fundamental level of full superstring theory, for example,
such a discrimination is not possible and one would therefore
not expect any decoherence effect to occur at that level.

In general one would expect not only one semiclassical
component of the form (\ref{semi}), but also many superpositions of such
terms. Since (\ref{wdw}) is a real equation, one would in particular
expect to have a superposition of (\ref{semi}) with its complex conjugate.
The no-boundary state in quantum cosmology has, for example,
such a form. Decoherence also acts between such semiclassical
branches, although somewhat less effective than within
one branch (Barvinsky et al.~1999c). For a macroscopic Universe,
this effect is big enough to warrant the consideration of
only one semiclassical component of the form (\ref{semi}).
This constitutes a symmetry-breaking effect similar to the
symmetry breaking for chiral molecules: While in the former case
the symmetry with respect to complex conjugation is broken,
in the latter case one has a breaking of parity invariance
(compare Figures~2 and 3 above).

It is clear that decoherence can only act if there is
a peculiar, low-entropy, state for the very early Universe.
This lies at the heart of the arrow of time in
the Universe. 
A simple initial condition like the one in Conradi and Zeh (1991)
can in principle lead to a quantum state describing 
the arrow of time, see also Zeh (1999).

\subsection{Classicality of primordial fluctuations}
According to the inflationary scenario of the early Universe,
all structure in the Universe (galaxies, clusters of galaxies)
arises from {\em quantum} fluctuations of scalar fields and
scalar fluctuations of the metric. Because also fluctuations
of the metric are involved, this constitutes an effect
of (linear) quantum gravity. 

These early fluctuations manifest themselves as anisotropies
in the cosmic microwave background radiation and have been
observed both by the COBE satellite and earth-based telescopes.
Certainly, these observed fluctuations are classical stochastic
quantities. How do the quantum fluctuations become classical?

It is clear that for the purpose of this discussion
the global gravitational degrees of freedom can already
by considered as classical, i.e. the decoherence process
of Sect.~4.2 has already been effective. The role of the gravitational
field is then twofold: firstly, the expanding Universe
influences the dynamics of the quantum fluctuations.
Secondly, linear fluctuations of the gravitational field
are themselves part of the quantum system.

The physical wavelength of a mode with wavenumber $k$
is given by
\be \lambda_{phys}=\frac{2\pi a}{k}\ . 
\ee
Since during the inflationary expansion the Hubble parameter $H$
remains constant, the physical wavelength of the modes
leaves the particle horizon, given by $H^{-1}$, at a certain
stage of inflation, provided that inflation does not
end before this happens. Modes that are outside the 
horizon thus obey
\be \frac{k}{aH}\ll 1\ . \ee
It turns out that the dynamical behaviour of these modes
lies at the heart of structure formation. These modes re-enter
the horizon in the radiation-and matter-dominated phases
which take place after inflation. 

For a quantitative treatment, the Schr\"odinger equation (\ref{sch})
has to be solved for the fluctuations in the inflationary
Universe. The easiest example, which nevertheless exhibits the
same features as a realistic model, is a massless scalar field.
It is, moreover, most convenient to go to Fourier space and to
multiply the corresponding variable with $a$. The resulting
fluctuation variable is called $y_k$, see Kiefer and Polarski (1998)
for details. Taking as a natural initial state the ``vacuum state'',
the solution of the Schr\"odinger equation (\ref{sch}) for the
(complex) variables $y_k$ reads\footnote{Since there is no
 self-interaction of the field,
different modes $y_k$ decouple, which is why I shall suppress
the index $k$ in the following.}
\be
\chi(y,t) = \left(\frac{1}{\pi|f|^2}\right)^{1/2}
  \exp\left(-\frac{1-2\mbox{i}F}{2|f|^2}|y|^2\right)\; , \lb{gauss} 
\ee
where
\bea
|f|^2 &=& (2k)^{-1}(\cosh 2r+\cos 2\varphi\sinh 2r), \\
F &=& \frac{1}{2} \sin 2\varphi\sinh 2r\ , 
\eea 
and explicit expressions can be given for the time-dependent
functions $r$ and $\varphi$.
The Gaussian state (\ref{gauss}) is nothing but a {\em squeezed state},
a state that is well known from quantum optics. The parameters
$r$ and $\varphi$ have the usual interpretation as
squeezing parameter and squeezing angle, respectively.
It turns out that during the inflationary expansion
$r\to\infty$, $|F|\gg 1$, and $\varphi\to 0$ (meaning here
a squeezing in momentum). In this limit, the state (\ref{gauss})
becomes also a WKB state par excellence. As a result of this extreme
squeezing, this state cannot be distinguished within
the given observational capabilities from a classical
stochastic process, as thought experiments demonstrate
(Kiefer and Polarski~1998, Kiefer et al.~1998a).
In the Heisenberg picture, the special properties of the state (\ref{gauss})
are reflected in the fact that the field operators commute
at different times, i.e.
\be
[\hat{y}(t_1),\hat{y}(t_2)]\approx 0\ . \lb{qnd}
\ee
(Kiefer et al.~1998b). In the language of quantum optics, 
this is the condition for a quantum-nondemolition
measurement: An observable obeying (\ref{qnd}) can repeatedly
be measured with great accuracy. It is important to
note that these properties remain valid after the modes
have reentered the horizon in the radiation-dominated
phase that follows inflation (Kiefer et al.~1998a).

As is well known, squeezed states are very sensitive to
interactions with other degrees of freedom 
(Giulini et al.~1996). Since such interactions are 
unavoidably present in the early Universe, the question arises
whether they would not spoil the above picture. 
However, most interactions invoke couplings in
field amplitude space (as opposed to field momentum space)
and therefore,
\be
[\hat{y},\hat{H}_{int}] \approx 0\ ,
\ee
where $\hat{H}_{int}$ denotes the interaction Hamiltonian.
The field amplitudes therefore become an excellent pointer basis:
This basis defines the classical property, and due to (\ref{qnd})
this property is conserved in time. The decoherence time
caused by $\hat{H}_{int}$ is very small in most cases.
Employing for the sake of simplicity a linear interaction
with a coupling constant $g$, one finds
for the decoherence time scale (Kiefer and Polarski~1998)
\be
t_D\approx \frac{\lambda_{phys}}{g\mbox{e}^r}\ . 
\ee
For modes that presently re-enter the horizon, one has
$\lambda_{phys}\approx 10^{28}\mbox{cm}$, $\mbox{e}^r
\approx 10^{50}$ and therefore
\be
t_D\approx 10^{-31}g^{-1}\mbox{sec}\ .
\ee
Unless $g$ is very small, decoherence acts on a very short
timescale. This conclusion is enforced if higher-order
interactions are taken into account.
It must be noted that the interaction of the field modes
with its ``environment'' is an {\em ideal} measurement --
the probabilities are unchanged and the main predictions
of the inflationary scenario remain the same (which manifest
themselves, for example, in the form of the anisotropy spectrum
of the cosmic microwave background). This would not be the case,
for example, if one concluded that {\em particle number}
instead of {\em field amplitude} would define the robust
classical property. Realistic models of the early Universe
must of course take into account complicated nonlinear
interactions, see e.g. Calzetta and Hu (1995) and Matacz (1997).
Although these models will affect the values of the decoherence
timescales, the conceptual conclusions drawn above will
remain unchanged.

The results of the last two subsections give rise to the
{\em hierarchy of classicality} (Kiefer and Joos~1999):
The global gravitational background degrees of freedom
are the first variables that assume classical properties.
They then provide the necessary condition for other
variables to exhibit classical behaviour, such as the
primordial fluctuations discussed here. These then serve as the
seeds for the classical structure of galaxies and clusters
of galaxies that are part of the observed Universe.  

\section{Acknowledgements}
I thank the organisers of this school for inviting
me to this interesting and stimulating meeting. I am 
grateful to the Institute of Advanced Study Berlin for its
kind hospitality during the writing of this contribution.
I also thank John Briggs, Erich Joos, Alexander Kamenshchik,
Jorma Louko for their comments on this manuscript.
%

\clearpage
\addcontentsline{toc}{section}{Index}
\flushbottom
\printindex


\begin{thebibliography}{99}
 %
\addcontentsline{toc}{section}{References}
%
\bibitem{}{}{}
Al'tshuler, B.L., Barvinsky, A.O. (1996):
Quantum cosmology and physics of transitions with a change
of the spacetime signature. Physics-Uspekhi {\bf 39}, 429--459
%
\bibitem{}{}{}
Ashtekar, A. (1999): Quantum mechanics of geometry.
 Electronic report gr-qc/9901023
%
\bibitem{}{}{}
Barbour, J.B. (1997): Nows are all we need. In: {\em Time,
Temporality, Now}, edited by H.~Atmanspacher and E.~Ruhnau
(Springer, Berlin)
%
\bibitem{}{}{}
Barvinsky, A.O., Kamenshchik, A.Yu. (1994):
Quantum scale of inflation and particle physics of
the early Universe. Phys. Lett.~B {\bf 332}, 270--276
%
\bibitem{}{}{}
Barvinsky, A.O., Kamenshchik, A.Yu., Kiefer, C. (1999a):
Effective action and decoherence by fermions in quantum cosmology.
Nucl. Phys.~B {\bf 552}, 420--444
%
\bibitem{}{}{}
Barvinsky, A.O., Kamenshchik, A.Yu., Kiefer, C. (1999b):
Origin of the inflationary Universe.
Mod. Phys. Lett. A {\bf 14}, 1083--1088
%
\bibitem{}{}{}
Barvinsky, A.O., Kamenshchik, A.Yu., Kiefer, C., Mishakov, I.V.
(1999c): Decoherence in quantum cosmology at the onset of inflation.
Nucl. Phys.~B {\bf 551}, 374--396
%
\bibitem{}{}{}
Barvinsky, A.O., Kiefer, C. (1998): Wheeler-DeWitt equation
 and Feynman diagrams. Nucl. Phys. B {\bf 526}, 509--539
%
%
\bibitem{}{}{}
Blanchard, P., Giulini, D., Joos, E., Kiefer, C., 
Stamatescu, I.-O., eds. (1999): {\em Decoherence: Theoretical,
experimental and conceptual problems} (Springer, Berlin)
%
\bibitem{}{}{}
Bousso, R., Hawking, S.W. (1996): Pair creation of black holes
during inflation. Phys. Rev. D {\bf 54}, 6312--6322
%
\bibitem{}{}{}
Briggs, J.S., Rost, J.M. (1999): Time dependence in quantum
mechanics. Electronic report quant-ph/9902035
%
\bibitem{}{}{}
Brout, R., Parentani, R. (1999): Time in cosmology.
Int. J. Mod. Phys.~D {\bf 8}, 1--22
%
\bibitem{}{}{}
Calzetta, E., Hu, B.L. (1995): Quantum fluctuations, decoherence
of the mean field, and structure formation in the
early Universe. Phys. Rev. D {\bf 52}, 6770--6788
%
\bibitem{}{}{}
Coleman, S. (1988): Why there is nothing rather than something:
A theory of the cosmological constant. Nucl. Phys. B~{\bf 310},
643--668
%
\bibitem{}{}{}
Conradi, H.D., Zeh, H.D. (1991):
Quantum cosmology as an initial value problem.
Phys. Lett.~A~{\bf 154}, 321--326
%
\bibitem{}{}{}
D\c{a}browski, M.P., Kiefer, C. (1997): 
Boundary conditions in quantum string cosmology.
Phys. Lett.~B~{\bf 397}, 185--192
%
\bibitem{}{}{}
Giulini, D. (1995): What is the geometry of superspace?
Phys. Rev. D {\bf 51}, 5630--5635
%
\bibitem{}{}{}
 Giulini, D., Joos, E., Kiefer, C., Kupsch, J.,
Stamatescu, I.-O., Zeh, H.D. (1996): {\it Decoherence and the
Appearance of a Classical World in Quantum Theory} (Springer, Berlin)
%
\bibitem{}{}{}
Giulini, D., Kiefer, C. (1994): Wheeler-DeWitt metric and the
attractivity of gravity. Phy. Lett. A {\bf 193}, 21--24
%
\bibitem{}{}{}
Halliwell, J.J. (1991): Introductory lectures on quantum cosmology.
In {\it Quantum cosmology and baby universes}, edited by S. Coleman,
J.B. Hartle, T. Piran, and S. Weinberg (World Scientific, Singapore)
%
\bibitem{}{}{}
Haroche, S. (1998): Entanglement, decoherence and the
 quantum/classical boundary. Phys. Today {\bf 51} (July),
 36--42
%
\bibitem{}{}{}
Hartle, J.B. (1997): Quantum cosmology: problems for the
 21st century. In: {\em Physics in the 21st century}, edited by
 K. Kikkawa, H. Kunimoto, H. Ohtsubo (World Scientific, Singapore)
%
\bibitem{}{}{}
Hartle, J.B., Hawking, S.W. (1983): Wave function of the Universe.
Phys. Rev. D {\bf 28}, 2960--2975
%
\bibitem{}{}{}
Hawking, S.W., Turok, N.G. (1998): Open inflation without
false vacua. Phys. Lett. B {\bf 425}, 25--32
%
\bibitem{}{}{}
 Isham, C.J. (1992): Canonical quantum gravity and the problem
  of time. In:
{\em Integrable Systems, Quantum Groups,
 and Quantum Field Theories}, edited by L. A. Ibart and
M. A. Rodrigues (Kluwer, Dordrecht)
%
\bibitem{}{}{}
Joos, E. (1986): Why do we observe a classical spacetime?
Phys. Lett. A {\bf 116}, 6--8
%
\bibitem{}{}{}
Joos, E., Zeh, H.D. (1985): The emergence of classical properties
through interaction with the environment. Z. Phys.~B {\bf 59}, 223--243
%
\bibitem{}{}{} 
Kiefer, C. (1987): Continuous measurement of mini-superspace variables
by higher multipoles. Class. Quantum Grav. {\bf 4}, 1369--1382
%
\bibitem{}{}{}
Kiefer, C. (1990): Wave packets in quantum cosmology and the
cosmological constant. Nucl. Phys. B~{\bf 341}, 273--293
%
\bibitem{}{}{}
Kiefer, C. (1991): On the meaning of path integrals in
quantum cosmology. Ann. Phys. (N.Y.) {\bf 207}, 53--70
%
\bibitem{}{}{}
Kiefer, C. (1992): Decoherence in quantum electrodynamics
and quantum cosmology. Phys. Rev. D {\bf 46}, 1658--1670
%
\bibitem{}{}{}
Kiefer, C. (1993): Topology, decoherence, and semiclassical gravity.
Phys. Rev. D {\bf 47}, 5414--5421
%
\bibitem{}{}{}
Kiefer, C. (1994): The semiclassical approximation to quantum gravity.
In: {\em Canonical gravity: From Classical to Quantum}, edited
    by J.~Ehlers and H.~Friedrich (Springer, Berlin)
%
\bibitem{}{}{}
Kiefer, C. (1997): Does time exist at the most fundamental level?
In: {\it Time, Temporality, Now}, edited by H. Atmanspacher
and E. Ruhnau (Springer, Berlin)
%
\bibitem{}{}{}
Kiefer, C. (1999): Decoherence in situations involving
the gravitational field. In: Blanchard et al. (1999).
%
\bibitem{}{}{}
Kiefer, C., Joos, E. (1999): Decoherence: Concepts and examples.
In: {\it Quantum Future}, edited by P.~Blanchard and A.~Jadczyk
(Springer, Berlin)
%
\bibitem{}{}{}
Kiefer, C., Lesgourgues, J., Polarski, D., Starobinsky, A.A. (1998a):
The coherence of primordial fluctuations produced during inflation.
Class. Quantum Grav. {\bf 15}, L67--L72
%
\bibitem{}{}{}
Kiefer, C., Polarski, D. (1998): Emergence of classicality for
 primordial fluctuations: concepts and analogies.
Ann. Phys. (Leipzig) {\bf 7}, 137--158
%
\bibitem{}{}{}
Kiefer, C., Polarski, D., Starobinsky, A.A. (1998b): 
Quantum-to-classical transition for fluctuations in the early universe.
Int. J. Mod. Phys. D {\bf 7}, 455--462
%
\bibitem{}{}{}
Kiefer, C., Singh, T.P. (1991): Quantum gravitational corrections
to the functional Schr\"odinger equation.
Phys. Rev. D~{\bf 44}, 1067--1076
%
\bibitem{}{}{}
Kiefer, C., Zeh, H.D. (1995): Arrow of time in a recollapsing
quantum universe. Phys. Rev. D~{\bf 51}, 4145--4153
%
\bibitem{}{}{}
 Kucha\v{r}, K.V. (1992): Time and interpretations of quantum gravity.
 In: {\em Proceedings of the 4th Canadian Conference
  on General Relativity and Relativistic Astrophysics},
  edited by G. Kunstatter, D. Vincent and J. Williams
  (World Scientific, Singapore)
%
\bibitem{}{}{}
Lema\^{\i}tre, G. (1931): The beginning of the world from the
point of view of quantum theory. Nature {\bf 127}, 706
%
\bibitem{}{}{}
Matacz, A. (1997): A new theory of stochastic inflation.
Phys. Rev. D {\bf 55}, 1860--1874
%
\bibitem{}{}{}
Mott, N.F. (1931): On the theory of excitation by collision
with heavy particles. Proc. Cambridge Phil. Soc. {\bf 27}, 553--560
%
\bibitem{}{}{}
Rosales, J.-L. (1997): Quantum state correction of relic gravitons
from quantum gravity. Phys. Rev. D~{\bf 55}, 4791--4794
%
\bibitem{}{}{}
Schr\"odinger, E. (1935): Discussion of probability relations
 between separated systems. Proc. Cambridge Phil. Soc. {\bf 31}, 555--563
%
\bibitem{}{}{}
Sen, A. (1998): Developments in superstring theory.
To be published in the Proceedings of the 29th International
Conference on High-Energy Physics, Vancouver, Canada,
 23-29 July~1998, 
 Electronic report hep-ph/9810356
%
\bibitem{}{}{}
Teitelboim, C. (1980): The Hamiltonian structure of space-time.
In: {\em General relativity and gravitation}, edited by A. Held (Plenum Press,
New York)
%
\bibitem{}{}{}
Vilenkin, A. (1998): The quantum cosmology debate.
Contribution to the conference on particle physics and the early
universe (COSMO 98), Monterey, CA, 15.-20.11.1998,
Electronic report gr-qc/9812027
%
\bibitem{}{}{}
Zeh, H.D. (1986): Emergence of classical time from a universal
wave function. Phys. Lett. A {\bf 116}, 9--12
%
\bibitem{}{}{}
Zeh, H.D. (1997): What is achieved by decoherence? In {\it
New Developments on Fundamental Problems in Quantum Physics}, edited by
M.~Ferrer and A.~van der Merwe (Kluwer Academic, Dordrecht)
%
\bibitem{}{}{}
Zeh, H.D. (1999): {\it The physical basis of the direction of time}
                  (Springer, Berlin)
%
\bibitem{}{}{}
 Zurek, W.H. (1991): Decoherence and the Transition
     from Quantum to Classical. Physics Today {\bf 44} (Oct.),
     36--44; see also the discussion in Physics Today (letters)
     {\bf 46} (April), 13
%

\end{thebibliography}
\end{document}